\def\Eq#1{Eq.~(\ref{#1})}
\def\dslash{\not\!\partial}
\def\aslash{\not \!\! A}
\def\jpialpha{J^{\alpha}_{\pi}}
\def\jpi{J_{\pi}}
\def\jsigma{J_{\sigma}}
\def\W{{\mathcal{W}}}

\def\N{\mathcal{N}}
\def\trf{\rm{Tr_{f}}}
\def\trc{\rm{Tr_{c}}}
\documentclass[12pt]{article}
\usepackage[]{}

\begin{document}

 \def\thepage{}
\begin{titlepage}
\title{Chiral Symmetry and Meson Vertex Operators 
in QCD Strings}
\author{\\
Vikram Vyas\footnote{Associate, Abdus Salam International
Center for Theoretical Physics, Trieste, Italy.}\\
{  {\textit{Scientific Resource Centre } } }\\
{\textit{The Ajit Foundation, Jaipur, India}} \\
{{{Email: visquare@satyam.net.in}}}
}
\date{}
\maketitle
\begin{abstract}
    The worldline representation of the one loop fermionic 
    effective action is used to obtain the vertex operator for the 
    pion and the sigma in QCD strings. The vertex operator of the 
    scalar sigma is distinguished
    from that of the pseudo-scalar pion by the presence of an 
    additional  operator $m V_{0}$ where $m$ is the current 
    quark mass and $V_{0}$ is a vertex operator that would describe a 
    tachyon in the open bosonic string theory. This 
    leads to a relation between the sigma propagator and the pion 
    propagator,  when chiral symmetry is spontaneously broken 
    this relation implies that the propagator constructed from 
    $V_{0}$ must behave like a massless ghost state. 
    The presence of this state ensures that the sigma is massive 
    and no longer degenerate with the pion.
    The expectation value of $V_{0}$ in a string description 
    is related to the vacuum expectation value of the chiral condensate
    in QCD. Our analysis emphasizes the need for boundary fermions in
    any string representation of mesons and is suggestive of 
    world sheet supersymmetry.
\end{abstract}

\end{titlepage}

\renewcommand{\thepage}{\arabic{page}}
\setcounter{page}{1}

\section{Introduction}\label{sec:1}

The large $N$ limit of QCD, where $N$ is the number of colors, has 
provided various useful insights as to how QCD describes the strong 
interactions~\cite{tHooft2}. It has also suggested the possibility 
of a string description of QCD
\cite{PolyBook, Polchinski92}.
In this letter we will further explore this possibility using the 
worldline  representation of the one loop fermionic effective action
(FEA)~\cite{schubert1, DHoker}.  
The worldline representation of the one 
loop FEA has largely been used as a tool for simplifying perturbative 
calculation of one loop diagrams  in the presence of external
fields~\cite{schubert}.  Here we will use the worldline representation 
as a bridge between the large $N$ QCD and its string representation. 

Starting from the worldline formalism of the real part of the 
one loop FEA we will construct vertex operators 
appropriate for describing meson propagators in the corresponding 
string representation.  An important feature of these vertex 
operators is that they explicitly take into account the spin degrees of 
freedom of the quarks and correspondingly involve worldline fermions. 

In developing a string description of the strong interactions a 
natural question to ask is how the spontaneous breaking of chiral 
symmetry is reflected in this representation. This classic question 
has been reviewed in~\cite{lewellen}. 
We will study it in the context of QCD strings and ask how the 
sigma and the pion vertex operator behave when chiral symmetry is 
spontaneously broken. We will find that what distinguishes the sigma 
vertex operator from the pion vertex operator is the presence of an 
additional term $m V_{0}$, where $m$ is the current quark mass and 
$V_{0}$ is a vertex operator that would describe a tachyon in the 
open Nambu-Goto bosonic string theory. This leads to a relation 
between the sigma and the pion propagator which in turn requires 
that when chiral symmetry is spontaneously broken the propagator 
for $m V_{0}$ must describe, unlike in the case of 
the Nambu-Goto strings, a massless ghost state. The presence of this 
ghost state ensures that the sigma is massive and no longer 
degenerate with the pion. It is worth noting that $V_{0}$ by itself 
does not correspond to any physical particle of the theory, but 
it's expectation value is related to the chiral condensate of the 
quark anti-quark pairs in QCD vacuum.

The outline of the paper is as follows. In the next section
we will remind ourselves how in the large $N$ limit the connected 
gauge invariant Green's functions can be written using the  
functional average over the gauge fields of the one loop FEA.  
We will also in this section summarize in our notation the  worldline 
representation of the real part of one loop FEA obtained 
in~\cite{schubert1, DHoker}. 
In section~3  we will construct meson vertex operators 
for describing processes involving even powers of pion 
field in the large $N$ limit of the theory. We will observe 
that these vertex operators are also appropriate for describing 
meson propagators in the string representation of the large $N$ QCD. 
In section~4 we will study how the spontaneous breaking of chiral 
symmetry is reflected in the behavior of the sigma and the 
pion vertex operator. We state our conclusions in the final section.

\section{The Large $N$ Limit and the Worldline\\
  Formalism}\label{sec:2}

We start with the familiar observation that in the large $N$ limit of 
QCD the gauge invariant Green's functions of quark bilinears can be 
obtained from the one loop FEA averaged over the gauge 
fields~\cite{WittenCarges, WitteN}.   Consider 
the Euclidean partition function for two quark flavors in the presence 
of a source for the sigma, $\jsigma$, and a source 
for the pion triplet, $\jpialpha$,
\begin{equation}
    Z[\jpi, \jsigma]  =  \int DA D\bar{\Psi}D\Psi \exp\{-S_{YM}[A]\} 
     \exp \left \{-\int_{x} \bar{\Psi}{O}[ A, \jpi, J_{\sigma} ] \Psi 
     \right \},
    \label{zpi}  
\end{equation}
$S_{YM}[A]$ represents the Yang-Mills action for the $SU(N)$ 
gauge field $A$ while $\Psi$ and $\bar{\Psi}$ are the quark fields.
The fermionic part of the action, using the Feynman 
slash notation, is given by
\begin{equation}
     {O}[A, \jpi, \jsigma]_{ij}^{ab}= -i(\!\!\dslash +m)\delta_{ab} \delta_{ij} 
       - \aslash_{ab}(x)\delta_{ij} - i\delta_{ab} 
     \delta_{ij}\jsigma -
     \delta_{ab} \gamma_{5} J^{\alpha}_{\pi}(x) \tau^{\alpha}_{ij},
    \label{eq:sfpi}
\end{equation}
where $a, b$ denotes the color indices while $i, j$ denotes the 
flavor indices, and $m$ is a flavor independent current quark mass. 
The generators of the flavor group  
$SU_{f}(2)$  are represented by $\tau^{\alpha}$ and we have
defined
\begin{equation}
    \jpi = \jpialpha \tau^{\alpha}.
    \label{defjpi}
\end{equation} 
A formal integration over the quark fields leads to
\begin{equation}
    Z[\jpi, \jsigma] = Z_{YM} < \exp\{-\Gamma[A, J_{\pi}, J_{\sigma}] \}>_{A},
    \label{zFEA}
\end{equation}
where $Z_{YM}$ is the partition function for the $SU(N)$ gauge theory in 
the absence of matter fields. $\Gamma[A, J_{\pi}, J_{\sigma}]$ is the 
one loop FEA.  The averaging over the gauge fields is defined by
\begin{equation}
     < \exp\{-\Gamma[A, \jpi, \jsigma] \}>_{A} =  \frac{1}{Z_{YM}}\int DA 
     \exp\{-S_{YM}[A]\}  \exp\{-\Gamma[A, J_{\pi}, J_{\sigma}] \}.
    \label{avgA}
\end{equation}
We define the connected pseudoscalar and scalar   
Green's functions as
\begin{eqnarray}
     \Delta^{\alpha\beta}_{\pi}(x-y) & = & \left ( \frac{\delta}{\delta 
    J^{\alpha}_{\pi}(x)} \frac{\delta}{\delta 
    J^{\beta}_{\pi}(y) } 
    < -\Gamma[A, J_{\pi}, J_{\sigma}] >_{A}\right )_{J_{\pi}=0; 
    J_{\sigma}=0},
    \label{pionprop}  \\
     \Delta_{\sigma}(x-y) & = & \left ( \frac{\delta}{\delta 
    J_{\sigma}(x)} \frac{\delta}{\delta 
    J_{\sigma}(y) } 
    < -\Gamma[A, J_{\pi}, J_{\sigma}] >_{A}\right )_{J_{\pi}=0; 
    J_{\sigma}=0}.
    \label{sigmaprop}
\end{eqnarray}
We will refer to above Green's functions as propagators implying that 
the long  distance behavior of these two-point connected Green's 
functions are governed by the corresponding  meson poles. 

The worldline formalism expresses  the
one loop FEA as a sum over closed paths of a spin-half particle. 
A worldline path integral for the one loop FEA in the presence of 
a scalar and a pseudoscalar source has been derived in 
\cite{schubert1, DHoker}. In the rest of this section we will 
restate their result in a notation that will clarify the link 
between the worldline  path integral  and a 
possible string representation of mesons in the large $N$ 
QCD.  

For the task at hand which involves Green's functions with even 
powers of $\gamma_{5}$ we will only 
need the worldline path integral for the real part of the one loop 
FEA~\cite{DHoker} which we will denote by 
$\Gamma_{R}[A, J_{\pi}, J_{\sigma}]$. 
It's functional average over the gauge fields can be written as
\begin{eqnarray}
    <-\Gamma_{R}[A, J_{\pi}, J_{\sigma}]>_{A} & = &
     \int_{0}^{\infty}\frac{dT}{T}  \exp\{ -m^{2}\frac{T}{2} \} 
     \nonumber \\
     &&\times <\trf [\hat{P} \exp \{-S[J_{\pi}, J_{\sigma}] \}] >_{x, 
     \psi_{\mu}, \psi_{5}, \psi_{6}}
    \label{wlineG}
\end{eqnarray}
where $\trf$ represents the 
trace over the flavor degrees of freedom while $\hat{P}$ is 
the path ordering operator for matrices, 
and the worldline  average for an arbitrary functional 
$F[x, \psi_{\mu}, \psi_{5}, \psi_{6}]$ is defined by the path 
integral
\begin{eqnarray}
      <F[x, \psi_{\mu}, \psi_{5}, \psi_{6}] >
      _{x, \psi_{\mu}, \psi_{5}, \psi_{6}} 
      & = &  
    -\frac{1}{4} \N[T] 
    \int D x  D \psi_{\mu} D \psi_{5}  D \psi_{6}\exp \{-S_{0}\} 
    \nonumber \\
    && \times <\W[x, \psi_{\mu}; A]>_{A}F[x, \psi_{\mu}, \psi_{5}, 
    \psi_{6}].
    \label{wlineAvg}  
\end{eqnarray}
The worldline path integral can be thought of as sum over closed 
paths of length $T$. A  path being described by the bosonic 
coordinates $x_{\mu}(\tau)$ 
and by a set of fermionic coordinates $\psi_{\mu}({\tau}), 
\psi_{5}(\tau), \psi_{6}(\tau)$ where
$\tau$ parametrizes the worldline and the index $\mu$ takes 
values from 1 to 4. The bosonic coordinates satisfy 
periodic boundary condition while the fermionic coordinates satisfy 
anti-periodic boundary condition. The functional integral over the 
fermionic coordinates is related to taking the trace over gamma 
matrices in the usual formalism~\cite{PolyBook}.
The action $S_{0}$ and the worldline action for the 
source term $S[J_{\pi}, J_{\sigma}]$ are given by
\begin{eqnarray}
    S_{0} & = & \int_{0}^{T} d\tau \{ \frac{\dot{x}^{2}}{2 } 
    +\frac{1}{2}\psi_{\mu}\dot{\psi_{\mu}} 
    +\frac{1}{2}\psi_{5}\dot{\psi_{5}} 
    + \frac{1}{2}\psi_{6}\dot{\psi_{6}}\},
    \label{s0}  \\
    S[J_{\pi}, J_{\sigma}] & = & \int_{0}^{T} d\tau \{ \frac{1}{2} 
    J_{\pi}^{2} + i  \psi_{\mu}\psi_{5}\partial_{\mu}J_{\pi} 
    + m J_{\sigma} 
     + \frac{1}{2} J_{\sigma}^{2} + 
    i  \psi_{\mu}\psi_{6}\partial_{\mu}J_{\sigma} 
    \},
    \label{sJpi}
\end{eqnarray}
a dot over a coordinate denotes a derivative with respect 
to the parameter $\tau$. The Wilson loop for a spin-half particle, 
$\W[x, \psi_{\mu}; A]$, 
which we will refer to as the fermionic Wilson loop, is defined by
 \begin{equation}
     \W[x, \psi_{\mu}; A]   =  \trc \hat{P} \exp \left \{ - i \int_{0}^{T} 
     d \tau \{ \dot{x}_{\mu} A_{\mu} - \frac{1}{2}\psi_{\mu}F_{\mu 
    \nu}\psi_{\nu} \} \right \},
     \label{fwl}
 \end{equation}
$\trc$ denotes the trace over the color indices, and 
$F_{\mu \nu}$ is the Yang-Mills field strength.  $\N[T]$ in 
(\ref{wlineAvg}) is a normalization factor for the functional 
integral and is given by
\begin{equation}
    \N[T]=\int Dp \exp \left \{ -\frac{1}{2} \int_{0}^{T} d \tau p^{2} 
    \right \}.
    \label{nT}
\end{equation}
In the next section we will use the above representation of the one 
loop FEA to describe mesons in the strings dual to the large $N$ QCD. 

\section{Vertex Operators for Mesons}\label{sec:3}

One natural way of stating the \ string - QCD duality is  
in terms of the Wilson Loop~\cite{PolyBook}
\begin{eqnarray}
    <W[x; A]>_{A} & = & 
    <\trc \hat{P} \exp \{i\oint A_{\mu}dx_{\mu}\}>_{A},
    \label{wloop}  \\
    <\trc \hat{P} \exp \{i\oint A_{\mu}dx_{\mu}\}>_{A}
    & = & \int D\Sigma \exp\{-S_{CS}[\Sigma, x]\},
    \label{ymstring}
\end{eqnarray}
where the functional integral is over the world-sheets whose 
boundary is the Wilson loop. The string action $S_{CS}$ is 
of course famously unknown, various attempts towards discovering it
have been 
reviewed in~\cite{Polyakov97, polyakov98}. 
One expects a similar string representation for the fermionic Wilson 
loop
\begin{equation}
    <\W[x,\psi; A]>_{A}=\int D\Sigma \exp\{-(S_{CS}[\Sigma, x] 
    +S_{B}[\Sigma, x, \psi] )\},
    \label{qcdstring}
\end{equation}
where $S_{B}[\Sigma, x, \psi] $ is a boundary action describing 
the interaction between the quark spin and the world-sheet degrees 
of freedom~\cite{vik2000}. Assuming this string-QCD duality
and using the worldline path integral for the one 
loop FEA allows us to obtain a string representation for the meson 
propagators.

Let us first consider the pion propagator, to obtain it's string 
representation we substitute the worldline path integral of the 
one loop FEA (\ref{wlineG}) in the definition of the pion propagator 
(\ref{pionprop}) to obtain
\begin{equation}
    \Delta^{\alpha \beta}_{\pi}(y_{1} - y_{2}) = \delta^{\alpha \beta}
    \int_{0}^{\infty} \frac{d T}{T} \exp \{-m^{2}  \frac{T}{2} \}
    <V_{5}(y_{1}) V_{5}(y_{2}) >_{x, \psi_{\mu}, \psi_{5}, \psi_{6}}
    \label{wlinePionProp}
\end{equation}
where the pion vertex function $V_{5}$ is given by 
\begin{eqnarray}
    \tau^{\alpha} V_{5}(y) &=& \left ( \frac{\delta}{\delta 
    J^{\alpha}_{\pi}(y)} \exp \{ -S[J_{\pi}, \jsigma] \} \right 
    )_{J_{\pi}=0; \jsigma=0},
    \label{defVpi} \\
     \tau^{\alpha} V_{5}(y) &=& -i \tau^{\alpha} \int_{0}^{T} 
    d\tau \left \{ \psi_{\mu}(\tau) \psi_{5}(\tau) \partial_{\mu}(
    \delta (x(\tau) - y) )\right \}.
    \label{Vpi}
\end{eqnarray}
In writing the pion propagator in terms of the vertex operator 
$V_{5}$ 
we have neglected a contact term that 
only contributes when $y_{1}$ and $y_{2}$ coincide, for we  
will be interested only in the relationship between the pion and the 
sigma propagator and an identical contact term will appear in the 
case of the sigma propagator too\footnote{
The contact term has the following form
$\int_{0}^{T}d\tau \delta (y_{1}-x(\tau)) \delta (y_{2}-x(\tau))$.}.
The vertex operator appropriate for describing sigma is 
given by 
\begin{equation}
    V_{\sigma}(y) = \left ( \frac{\delta}{\delta
    J_{\sigma}(y)} \exp \{ -S[\jpi, \jsigma] \} \right 
    )_{\jpi =0; \jsigma=0},
    \label{defVsigma}
\end{equation}
using (\ref{sJpi}) for $S[\jpi, \jsigma]$ leads to it's explicit form as
\begin{equation}
    V_{\sigma}(y) = m  V_{0}(y) + V_{6}(y),
    \label{Vsigma}
\end{equation}
where $V_{0}(y)$ and $V_{6}(y)$ are given by
\begin{eqnarray}
     V_{0}(y) & = & -  \int_{0}^{T} d \tau \delta ( x(\tau) - y ),
    \label{V0}  \\
    V_{6}(y) & = & -\int_{0}^{T} 
    d\tau \left \{ \psi_{\mu}(\tau) \psi_{6}(\tau) \partial_{\mu}(
    \delta (x(\tau) - y) )\right \}.
    \label{v6}
\end{eqnarray}
Again, as in the case of the pion, using the sigma vertex operator 
and substituting the worldline representation of the one loop FEA in 
the definition of the sigma propagator, \Eq{sigmaprop}, leads to  
\begin{equation}
    \Delta_{\sigma}(y_{1} - y_{2}) = 
    \int_{0}^{\infty} \frac{d T}{T} \exp \{-m^{2}  \frac{T}{2} \}
    <V_{\sigma}(y_{1}) V_{\sigma}(y_{2}) >_{x, \psi_{\mu}, \psi_{5}, 
    \psi_{6}},
    \label{wlineSigmaProp}
\end{equation}    
where we have again neglected the above mentioned contact term.

What we have been able to do above is to map the interpolating fields 
for mesons to geometrical quantities, the vertex operators, which are 
defined in terms of the quark worldline. Our expressions for meson 
propagators represents a formal sum of all the planer diagrams which
have one quark loop as their boundary. There has always been a hope 
that some string theory may provide a tractable way of summing the 
planer diagrams, \Eq{wlinePionProp} and \Eq{wlineSigmaProp}
expresses that possibility.  Importantly, using them we could 
identify the vertex operators that describe mesons in  
the unknown QCD string theory, 
but with an important caveat that 
the vertex operators so obtained can only  be used for 
processes involving  even number of pions in the leading 
approximation of the large $N$ limit. This is a consequence of the 
fact that they were derived from the real part of the one loop 
FEA. Though we will not need the imaginary part of the one loop 
FEA, it too can be written in the worldline formalism, 
but not in a unique manner~\cite{DHoker}. 

\section{Chiral Limit in the String Representation}\label{sec:4}

The idea that the pion is an approximate Nambu-Goldstone boson 
of a spontaneously broken chiral symmetry has proved to be an 
extremely useful one~\cite{NJL}. If the large $N$ limit is a good 
approximation to QCD then chiral symmetry should  be 
spontaneously broken for an arbitrary large value of $N$. A strong 
evidence in favor of this is provided by the Coleman-Witten 
theorem~\cite{coleman} which shows that under some reasonable 
assumptions not only  chiral symmetry is spontaneously broken in 
the large $N$ limit but also the pattern of breaking is the one that
is observed in nature.

We would like to see how the spontaneous breaking of chiral symmetry 
is reflected in the meson vertex operators that we obtained in the 
previous section.
For this purpose it will be convenient to work with 
the momentum space vertex operators and propagators and we will 
ignore the flavor index as it does not play any role in the leading term 
of the large $N$ expansion. 
Let us first consider the pion propagator in the chiral limit. 
The momentum space pion vertex operator is
\begin{equation}
    V_{5}(k) = \int_{0}^{T} d \tau \left \{ 
    k . \psi(\tau) \psi_{5}(\tau) \exp \{ i k . y(\tau)\}  \right \}
    \label{vpik},
\end{equation}
and the momentum space pion propagator is given by
\begin{equation}
    \Delta_{\pi}(k)  =  \int_{0}^{\infty} \frac{ d T}{T} \exp \{ 
    -m^{2}\frac{T}{2} \}
     < V_{5}(k) V_{5}(-k) >_{y, \psi_{\mu}, \psi_{5}, \psi_{6}}.
    \label{pikprop}
\end{equation}
In the worldline functional integral we have eliminated the zero 
mode from the path $x(\tau)$. The resulting path $y(\tau)$ satisfies
the condition
\begin{equation}
    \int_{0}^{T}d \tau y(\tau) = 0.
    \label{defytau}
\end{equation}
In the chiral limit the pion is a Nambu-Goldstone boson therefore it's
propagator will have the following form
\begin{equation}
    \lim_{m \rightarrow 0} \Delta_{\pi}(k) = \frac{F_{\pi}(k^{2})}{k^{2}},
    \label{ngboson}
\end{equation}
where the unknown function $F_{\pi}(k^{2})$ is regular at $k^{2} = 0$. 
The key point is that the pion 
propagator is well defined as the quark mass goes to zero and has a
pole at $k^{2}=0$.
Now let us consider the momentum space propagator  for the sigma 
in the chiral limit. The momentum space vertex operator for the sigma 
is given by 
\begin{equation}
    V_{\sigma}(k) = m V_{0}(k) + V_{6}(k),
    \label{vsigmak}
\end{equation}
where the vertex operator $V_{0}(k)$ and $V_{6}(k)$ are 
given by
\begin{eqnarray}
     V_{0}(k) & = & -\int_{0}^{T} d \tau \exp \{ i k . y(\tau) \},
    \label{v0k}  \\
    V_{6}(k) & = & \int_{0}^{T} d \tau \left \{ 
    k . \psi(\tau) \psi_{6}(\tau) \exp \{ i k . y(\tau)\}  \right \}.
    \label{v6k}
\end{eqnarray}
Using these operators we can write the sigma  propagator 
as
\begin{eqnarray}
    \Delta_{\sigma}(k) & = & \int_{0}^{\infty} \frac{ d T}{T} \exp \{ 
    -m^{2}\frac{T}{2} \}
     < V_{\sigma}(k) V_{\sigma}(-k) >_{y, \psi_{\mu}, \psi_{5}, 
     \psi_{6}}
    \label{eq:sigmakprop}  \\
     & = & m^{2}\Delta_{0}(k) + \Delta_{6}(k),
    \label{sigmakprop}
\end{eqnarray}
where we have defined $\Delta_{0}(k)$ and $\Delta_{6}(k)$ as
\begin{eqnarray}
     \Delta_{0}(k) & = & \int_{0}^{\infty} \frac{ d T}{T} \exp \{ 
    -m^{2}\frac{T}{2} \}
     < V_{0}(k) V_{0}(-k) >_{y, \psi_{\mu}, \psi_{5}, \psi_{6}}
    \label{tachprop}  \\
    \Delta_{6}(k) & = & \int_{0}^{\infty} \frac{ d T}{T} \exp \{ 
    -m^{2}\frac{T}{2} \}
     < V_{6}(k) V_{6}(-k) >_{y, \psi_{\mu}, \psi_{5}, \psi_{6}}
    \label{6prop}
\end{eqnarray}
In writing \Eq{sigmakprop} we have used the 
fact that the cross terms 
\begin{equation}
    <V_{6}(k) V_{0}(-k)>_{y, \psi_{\mu}, \psi_{5}, \psi_{6}} = 0
    \label{eq:crossterm}
\end{equation}
because the integrand is odd in $\psi_{6}(\tau)$. 
Next we notice that $\Delta_{\pi}(k)$ and $\Delta_{6}(k)$ are 
identical, for the right hand side of \Eq{6prop} is identical to the 
right hand side of \Eq{pikprop} apart from a  relabeling of the 
integration variable $\psi_{6}$ as $\psi_{5}$, thus
\begin{equation}
    \Delta_{\pi}(k) = \Delta_{6}(k).
    \label{eq:equality}
\end{equation}
This immediately gives us a relation  between the sigma and the 
pion propagator
\begin{equation}
    \Delta_{\sigma} (k) = m^{2} \Delta_{0}(k) + \Delta_{\pi}(k)
    \label{relation},
\end{equation}
before considering the above relation in the chiral limit, let us note 
that it would not be modified even if we had included the 
contact term in the pion and the sigma propagator (see the comments 
after~\Eq{Vpi}),
for the contact term for both the propagators are identical.
Let us now look at this relation in the chiral limit, it implies
that $\Delta_{0}(k)$ cannot have a smooth limit as the current 
quark mass $m$ goes to zero, otherwise the pion and the 
sigma will become 
degenerate and that will violate our assumption of the spontaneous 
breaking of chiral symmetry. In fact, we know from the 
phenomenology of the strong interactions that there is no evidence 
for a light scalar particle.
Thus, the propagator $\Delta_{0}(k)$ must have the following form
\begin{equation}
    \lim_{m \rightarrow 0} m^{2} \Delta_{0}(k) = 
    -\frac{F_{\pi}(0)}{k^{2}} + H(k^{2}),
    \label{tacghost}
\end{equation}
where the unknown function $H(k^{2})$ is regular at $k^{2}=0$. 
In other words  when the chiral symmetry is spontaneously broken 
the state corresponding to the vertex operator $V_{0}$  is
a massless ghost state. We note that $V_{0}$ by
itself does not correspond to any physical particle. It only appears as 
the part of the sigma vertex operator. We are now in a better position to understand~\Eq{relation} 
which seems to suggest that in the absence of 
quark mass the pion and sigma propagators are identical.
In particular it seems to
imply that the pion and the sigma have identical mass.
This indeed would be the case if chiral symmetry was not 
spontaneously broken (see for e.g~\cite{susskind}). 
If  chiral symmetry is spontaneously broken then one has to study 
the relation~(\ref{relation})
in the limit of the current quark mass $m$ going to zero. Our analysis 
revels that in this phase $m^{2} \Delta_{0}(k)$ does not vanish as 
$m$ 
tends to zero. The pion and the sigma are no longer degenerate. 
In fact one can readily see that the vacuum 
expectation value of the chiral condensate is given by the 
expectation value of the vertex operator $m V_{0}$,
\begin{equation}
    <\bar{\Psi}\Psi(k)> = \lim_{m \rightarrow 0}
    \delta (k)\int_{0}^{\infty} 
    \exp \{ -m^{2}\frac{T}{2} \}
    \frac{dT}{T}<-m V_{0}(k)>_{y, \psi_{\mu}, \psi_{5}, \psi_{6}}.
    \label{chiralcond}
\end{equation}
The above equation can be regarded as another statement of the 
large QCD string duality, the left hand side is the expectation 
value of the quark field operators while the right hand side is a 
worldline - string functional integral.

By the same account if the chiral symmetry was not spontaneously
broken then $m^{2} \Delta_{0}(k)$ should vanish in the limit of
$m$ going to zero and so would the expectation value of $m V_{0}$.  
The vertex operator $V_{0}$ is of course familiar in the open 
bosonic string theory where it describes a 
tachyon~\cite{stringbooks}. In the string theory dual to the large 
$N$ QCD it is unique in that it neither excites (boundary)
fermions nor the world sheet degrees of freedom, but its behavior 
allows us to delineate the correct ground state of the theory.

\section{Conclusions}\label{sec:5}

The large $N$ expansion of QCD has given surprisingly many 
unique insights as to how QCD describes the strong interactions. 
They are surprising because they are obtained without being able 
to calculate even the leading term of the large $N$ expansion.
In the present paper we have written these leading terms 
as a functional integral over the quark worldline together 
with suitable worldline vertex operators. Our motivation for this was 
to use the worldline formalism as an intermediate step towards the 
string description of the large $N$ QCD.  In doing so we found that 
the worldline vertex operators for mesons are also the vertex 
operators for mesons in their string representation, at least as 
far as describing processes involving even number of pion operators 
in the leading approximation of the large $N$ limit. 
An important feature of these vertex operators 
is that they involve the worldline or the boundary fermions.
Thus the string theory that we seek must have at the least
boundary fermions
and plausibly even world-sheet fermions 
with some kind of world-sheet supersymmetry as suggested by
Polyakov~\cite{Polyakov97}. The relationship between the one 
component boundary fermions and the two component world sheet 
fermions, together with the vertex operators that we have obtained, 
may perhaps give us some clue as to the nature of the QCD strings.

Using these vertex operators we were also able to
get some hint as to how the spontaneous breaking of
chiral symmetry is reflected in the string description of the large 
$N$ QCD.  In the spontaneously broken phase of the large $N$ QCD
the vertex operator which would have described a tachyon in the 
Nambu-Goto open strings, instead describes a massless ghost state 
while it's expectation value gives the chiral condensate.
Reassuringly, we also saw that this does not do any violence to the 
theory, this vertex operator by itself does not represent any particle 
of the theory but is a part of the vertex operator that 
describes the sigma. The occurrence of the ghost state simply 
ensures that the sigma is massive when chiral symmetry is 
spontaneously broken.  
There has always been an intuition that once the correct ground state 
of a string theory is identified then there should be no tachyons in it,
for the strings describing the large $N$ QCD this indeed seems to be
the case. 

\section*{Acknowledgments}

I am grateful to N. D. Hari Dass, Carleton~DeTar,  and 
Ansar~Fayyazuddin  for their  critical and very useful comments 
on a preliminary version of this paper. 
I would also like to thank K. S. Narain for very useful discussions 
and  Rajesh Gopakumar for pointing out an error in the earlier version.

\end{document}